\documentclass[12pt, preprint]{aastex}

\shorttitle{Six Transits of XO-2b}

\shortauthors{Fernandez et al.}

\newcommand\kms{\ifmmode{\rm km\thinspace s^{-1}}\else km\thinspace s$^{-1}$\fi}
\newcommand\ms{\ifmmode{\rm m\thinspace s^{-1}}\else m\thinspace s$^{-1}$\fi}
\newcommand\cmss{\ifmmode{\rm cm\thinspace s^{-2}}\else cm\thinspace s$^{-2}$\fi}
\newcommand\msun{\ifmmode{M_{\odot}}\else $M_{\odot}$\fi}
\newcommand\rsun{\ifmmode{R_{\odot}}\else $R_{\odot}$\fi}
\newcommand\mjup{\ifmmode{M_{\rm Jup}}\else $M_{\rm Jup}$\fi}
\newcommand\rjup{\ifmmode{R_{\rm Jup}}\else $R_{\rm Jup}$\fi}

\begin{document}

\bibliographystyle{apj}

\title{The Transit Light Curve Project.\\
 XII.~Six Transits of the Exoplanet XO-2b}

\author{Jose M.\ Fernandez\altaffilmark{1,2},
Matthew J.\ Holman\altaffilmark{1},
Joshua N.\ Winn\altaffilmark{3},
Guillermo Torres\altaffilmark{1},
Avi Shporer\altaffilmark{4},
Tsevi Mazeh\altaffilmark{4},
Gilbert A.\ Esquerdo\altaffilmark{1},
Mark E.\ Everett\altaffilmark{5}
}

\altaffiltext{1}{Harvard-Smithsonian Center for Astrophysics, 60
Garden Street, Cambridge, MA 02138, USA; \\
jfernand@cfa.harvard.edu.}

\altaffiltext{2}{Department of Astronomy, Pontificia Universidad
Cat\'{o}lica, Casilla 306, Santiago 22, Chile.}

\altaffiltext{3}{Department of Physics, and Kavli Institute for
Astrophysics and Space Research, Massachusetts Institute of
Technology, Cambridge, MA 02139, USA.}

\altaffiltext{4}{Wise Observatory, Raymond and Beverly Sackler Faculty
  of Exact Sciences, Tel Aviv University, Tel Aviv 69978, Israel.}

\altaffiltext{5}{Planetary Science Institute, 1700 East Fort Lowell
Road, Suite 106, Tucson, AZ 85719, USA.}

\begin{abstract}

We present photometry of six transits of the exoplanet XO-2b. By
combining the light-curve analysis with theoretical isochrones to
determine the stellar properties, we find the planetary radius to be
$0.996 ^{+0.031}_{-0.018} \ \rjup$ and the planetary mass to be $0.565
\pm 0.054 \ \mjup$. These results are consistent with those reported
previously, and are also consistent with theoretical models for gas
giant planets. The mid-transit times are accurate to within 1~min and
are consistent with a constant period. However, the period we derive
differs by 2.5$\sigma$ from the previously published period. More data
are needed to tell whether the period is actually variable (as it
would be in the presence of an additional body) or if the timing
errors have been underestimated.

\end{abstract}

\keywords{planetary systems---stars:~individual (XO-2, GSC
  03413-00005)}

\section{Introduction}

Observations of exoplanetary transits have provided the first
empirical information about the internal structure, composition,
surface temperature and atmospheric dynamics of planets outside the
Solar system \citep{2007prpl.conf..701C}. It is hoped that precise
measurements of transit times and durations will also provide a new
channel for the detection of low-mass planets
\citep{2005Sci...307.1288H, 2005MNRAS.359..567A}. Among the immense
amount of information these systems may deliver, the most fundamental
parameter is the radius of the planet. Accurate radius determination
is not a trivial task, and in some cases after a transiting planet has
been announced, more precise data have led to significant revisions on
their radius, with consequences for theories of planetary interiors
and atmospheres \citep[see,
e.g.][]{2008ApJ...683.1076W,2008arXiv0812.0029J}.

The Transit Light Curve (TLC) project is an effort to gather precise
photometry of exoplanetary transits, in order to determine the
fundamental properties of these planets and their parent stars as
accurately as possible, and to seek evidence of undetected planets in
the pattern of transit times and durations.

The present paper is concerned with the XO-2 system
\citep{2007ApJ...671.2115B}. This was the second system discovered by
the XO project \citep{2006ApJ...648.1228M}. The XO-2 star is a nearby
Solar-type star with a binary companion (identified through its common
proper motion), and has a high metallicity ($\left[{\rm
Fe/H}\right]=0.45$) compared to other planet-hosting stars. The XO-2b
planet is a normal hot Jupiter (if the word ``normal'' may be applied
to such interesting objects), with an orbital period $P = 2.615857 \
d$, mass $M = 0.57 \mjup$, and radius $R = 0.98 \rjup$
\citep{2007ApJ...671.2115B}.

In this work we present and analyze differential photometry of six
transits of XO-2b. We measured the mid-transit time for each of the
six events, and by combining the data, we obtained new and independent
estimates for the system parameters. This paper is organized as
follows. In $\S$~\ref{obs_red} we describe the observations and data
reduction procedure. In $\S$~\ref{model} we describe the model and the
techniques we used to estimate the physical parameters of this
system. In $\S$ \ref{results} we present the results for the planetary
parameters and mid-transit times. Finally, in $\S$~\ref{summ} we
present a brief summary of this work.

\section{Observations and Data Reduction}
\label{obs_red}

We observed four complete transits, and one partial transit, with the
1.2~m (48-inch) telescope at the Fred L.\ Whipple Observatory (FLWO)
on Mount Hopkins, in Arizona. The FLWO 1.2m telescope is equipped with
KeplerCam, a monolithic 4K$\times$4K Fairchild 486 CCD that gives a
$23 \arcmin \times 23 \arcmin$ field and a pixel size of $0.68
\arcsec$ in 2$\times$2 binned mode. We used a Sloan $z$ band filter to
minimize the effects of limb darkening on the shape of the transit
light curve.

We observed one complete transit with the Centurion 0.46~m (18-inch)
telescope, at the Wise Observatory, in Israel.  The Centurion
telescope is equipped with a Santa Barbara Instrument Group (SBIG)
ST-10 XME USB CCD camera. This thermoelectrically-cooled chip has 2184
$\times$ 1472 pixels each 6.8 $\mu$m wide, which convert to 1.1 arcsec
at the $f$/2.8 focus of the telescope. The chip offers a $40.5 \arcmin
\times 27.3 \arcmin$ field of view.  Observations from Wise were made
with no filter, because this camera has no filter wheel. The 18-inch
telescope CCD has an overall response similar to a ``wide-R'' band
\citep[for further details, see][]{2008Ap&SS.314..163B}.

\begin{figure}
\plotone{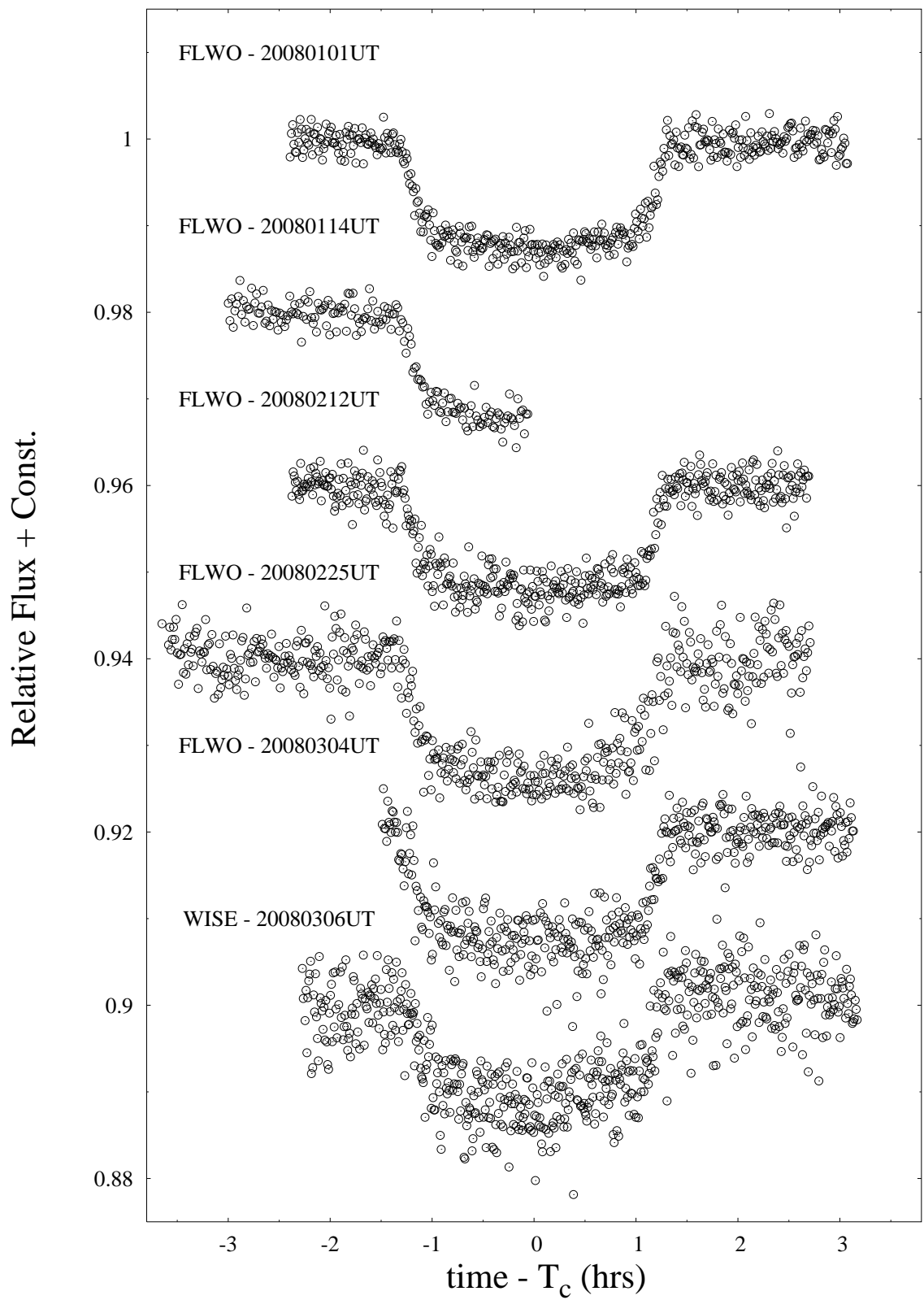}
\caption{Individual transit light curves, after applying a correction
  for differential extinction.  The FLWO transits were observed in the
  Sloan $z$ band. No filter was used at Wise; the effective bandpass
  resembles wide $R$ band. \label{indv_lc}}
\end{figure}

Relative aperture photometry was performed for XO-2 and 13 nearby
comparison stars, including the binary companion of XO-2 (7 stars were
used for the Wise data set). The choice of comparison stars was
iterative; we began with a longer list, and those stars that showed
unusual noise or variability were removed. The 13 comparison stars
range in brightness from 50--200\% of the brightness of XO-2. The sum
of the fluxes of the comparison stars was taken to be the reference
signal, and the flux of XO-2 was divided by this reference signal. The
root-mean-square (rms) noise in the relative flux of XO-2 among the
six light curves ranges between 0.0013 and 0.0036. Table
\ref{tab_phot} gives a summary of the six photometric observations,
with information about the dates, epochs, band, exposure times and
RMS. Figure \ref{indv_lc} shows the observed light curves. A
differential airmass correction was applied to the plotted light
curves. This correction was determined as part of the fitting process
that is described in the next section.  Table \ref{sample_lc} has an
abbreviated versions of one of the light curves. We intend for the
complete photometric data set to appear in the electronic version of
the journal.

\section{Data Modeling}
\label{model}

\subsection{Light Curve Analysis}

We modeled the light curves using the analytic formulas of
\citet{2002ApJ...580L.171M}. The orbital period was held fixed using
the value of Burke et al. (2007) and the planetary orbit was assumed
to be circular. The fitted parameters were the radius ratio
$R_{p}/R_{\star}$, the normalized semi-major axis $a/R_{\star}$, the
impact parameter $b$, the mid-transit time $T_c$, and two parameters
that take into account the zero-point and slope of the differential
airmass correction. We assumed a quadratic limb-darkening law, with
coefficients $u_1$ and $u_2$. Through a trial fit, we found that the
data are not precise enough to fit for both limb-darkening
coefficients. The linear combination $2u_{1}+u_{2}$ is
well-constrained, and the orthogonal combination $u_{1}-2u_{2}$ is
poorly constrained. For this reason we allowed $2u_{1}+u_{2}$ to be a
free parameter, and held $u_{1}-2u_{2}$ fixed.  We fixed
$u_{1}-2u_{2}$ at the value tabulated by
\citet{2004A&A...428.1001C}\footnote[1]{From
\citet{2004A&A...428.1001C}, $u_{1}=0.292$ and $u_{2}=0.299$}, for a
star with the metallicity, gravity, and temperature derived by
\citet{2007ApJ...671.2115B}.

The fitting statistic was
\begin{equation}
\label{chisq_lc}
\chi^{2}_{lc} = \sum_{j=1}^{N_{f}} \left[ \frac{f_{j}^{obs} - f_{j}^{mod}}{\sigma_{j}}\right]^{2}
\end{equation}
where $f_{j}^{obs}$ and $f_{j}^{mod}$ are the observed and modeled
relative fluxes observed at time $j$, and $\sigma_{j}$ is the
uncertainty of the data points as described below. To set the
appropriate data weights, we used a procedure that attempts to account
for any time-correlated (``red'') noise, at least approximately. For
each light curve we found an initial best-fitting model (using the
out-of-transit RMS as the flux weight) and calculated $\sigma_{1}$,
the standard deviation of the unbinned residuals between the observed
and calculated fluxes. Next we averaged the residuals into $M$ bins of
$N$ points and calculated the standard deviation of the binned
residuals. In the absence of red noise, we would have expected
\begin{equation}
\sigma_{N} = \frac{\sigma_{1}}{\sqrt{N}} \sqrt{\frac{M}{M-1}}
\end{equation}
but often the measured value of $\sigma_{N}$ is larger than this by a
factor $\beta$ due to time-correlated noise. In such cases the number
of effectively independent data points is smaller than the actual
number of data points. In Eq. \ref{chisq_lc} we used
$\sigma_{j}=\sigma_{1}\times\beta$. We found that $\beta$ depends
weakly on the choice of averaging time $\tau$, generally rising to an
asymptotic value at $\tau \sim 10$ minutes. The value $\beta$ was
close to unity except for the light curves of UT~2008~Feb~25
($\beta=1.2$) and UT~2008~Mar~06 ($\beta=1.5$).

The best-fit parameters and their uncertainties were obtained using a
Markov Chain Monte Carlo (MCMC) procedure. As described by
\citet{2005AJ....129.1706F} and \citet{2006ApJ...652.1715H}, in this
method a random process is used to create a sequence of points in the
parameter space that approximates the studied probability
distribution. This sequence or chain is generated by a jump function
that adds a Gaussian random number to each parameter. The jump is
executed if the new point has a $\chi^{2}_{lc}$ lower than the
previous point. If $\chi^{2}_{lc}$ is larger, the jump is made with a
probability equal to exp$\left(-\Delta\chi^{2}_{lc}/2\right)$. If the
jump is not made, the new point is a copy of the previous one. The
relative sizes of the perturbation were set equal to the approximate
$1\sigma$ uncertainties obtained by direct inspection of
$\chi^{2}_{lc}$ across the parameter space, as done in
\citet{2007ApJ...663..573B}. The sizes of the jumps are set by
requiring that $\sim$25\% of the jumps are accepted. As a first step,
four independent chains of $2 \times 10^5$ points were created for
each light-curve, discarding the first 20 \% of points to minimize
transient effects. The four chains were then superposed to create one
sequence of points.  The Gelman \& Rubin (1992) R statistic was always
within 0.2\% of unity for each parameter, a sign of good mixing and
convergence. The cumulative posterior probability is a normalized
histogram of the MCMC sample values. To derive the confidence interval
for a parameter, the MCMC sets were sorted by the parameter of
interest and then we determined the 15.85\%, 50\%, and 84.15\% points
of the cumulative posterior distribution.  The 50\% point (i.e.~the
median) was taken to be the ``best-fitting value'' and the interval
between the 15.85\% and 84.15\% points was taken to be the 68.3\%
(1--$\sigma$) confidence interval.

\begin{figure}
\plotone{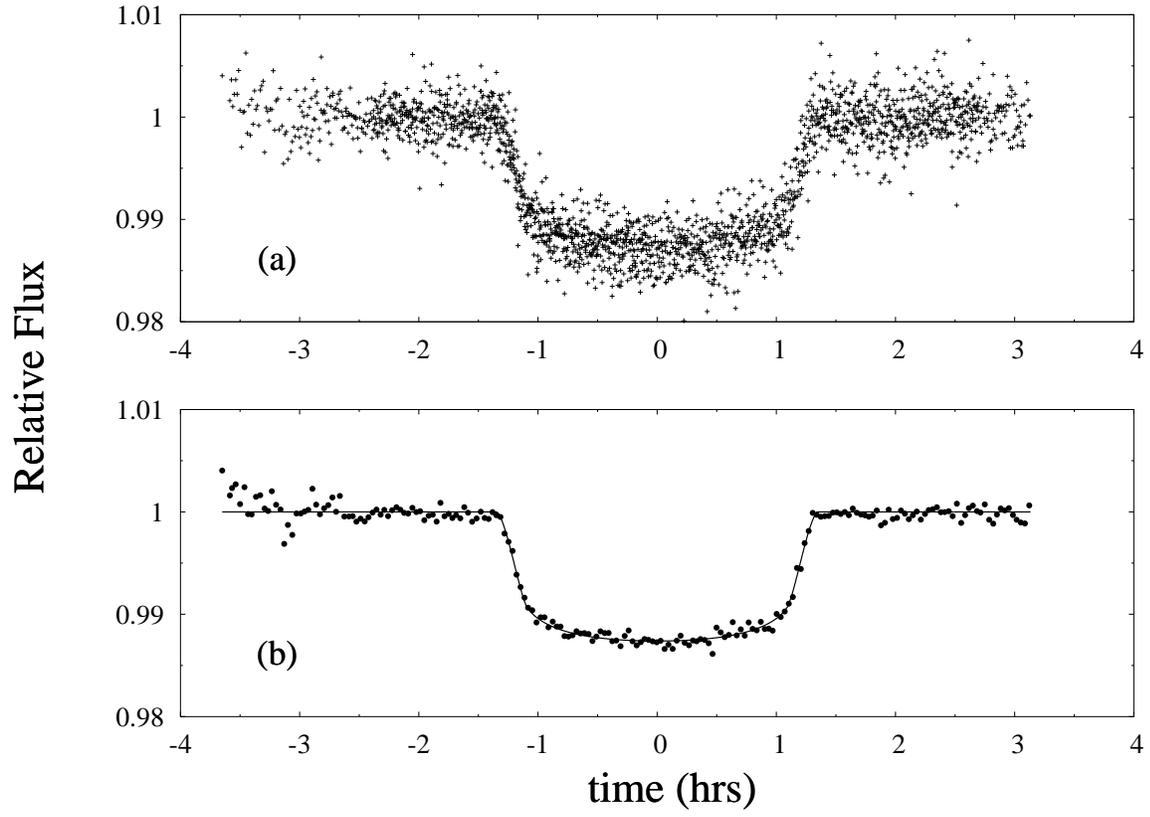}
\caption{A. Combined FLWO light curves. B. Best-fit model and 2 min
binned combined light curve. The light curve observed from Wise is not
included in this combined data set.
 \label{comb_lc}}
\end{figure}

The final values for the transit parameters $R_{p}/R_{\star}$,
$a/R_{\star}$ and $b$ were obtained by analyzing the combined data
from the five light curves observed with FLWO/KeplerCam, following the
same procedure used for the individual transits. A 2-minute binned
version of these combined light curves is shown in Figure
\ref{comb_lc}.

For the transit timings, we also performed two independent additional
tests to check the accuracy of our results. For one of the tests we
used a standard bootstrap simulation: we created $10^4$
``realizations'' of the data set by perturbing the best-fitting model
with Gaussian random noise (with a standard deviation equal to the rms
of the actual light curve), and minimizing $\chi^2_{lc}$ as a function
of $t_{c}$ for each realization ($R_{p}/R_{\star}$, $a/R_{\star}$ and
$b$ are not correlated with $t_{c}$, and were therefore held fixed for
these tests). The resulting collection of $10^4$ timings for each
transit was taken to be the probability distribution of the timing
\citep[see, e.g.,][]{1992nrfa.book.....P}. For the second test we
implemented the residual-permutation or ``rosary-bead'' method. For
each light curve, we found the best-fitting model and computed the
time series of $N$ residuals. We then added these residuals to the
model light curve after performing a cyclic permutation of the time
indices. This is a variant of the bootstrap technique that preserves
the temporal correlations among the residuals, and has been used
previously by many investigators \citep[e.g.,][]{2005A&A...431.1105B,
2008MNRAS.386.1644S}. For the six light curves under study, the
timings obtained on the additional tests were consistent with our
first set of results, and with similar error bars. The results of the
light curve analysis are summarized in Table \ref{tab_params}.

\subsection{Determination of Absolute Dimensions}

As has been shown before by several authors, the only intrinsic
properties of the star and planet that can be determined directly from
observed quantities on transiting systems are the mean density of the
star \citep{2003ApJ...585.1038S} and the surface gravity of the planet
\citep{2007ApJ...664.1190S, 2007ApJ...664.1185H,
  2004MNRAS.355..986S}. In order to determine the individual masses
and radii of the bodies, external information must be introduced.

In this case, we used stellar evolution models following the procedure
and considerations of \citet{2008ApJ...677.1324T}. For this purpose we
rely on the spectroscopic temperature ($T_{\rm eff}=5340\ K$) and
metallicity ($\left[{\rm Fe/H}\right]=0.45$) obtained by
\citet{2007ApJ...671.2115B}, but we adopt more conservative errors, no
smaller than 0.05 dex in $\left[{\rm Fe/H}\right]$ and 80 K in
$T_{\rm eff}$. The reason for this approach is to take into account the
documented difficulty in obtaining accurate values for stellar
temperatures \citep[see, eg.,][]{1977MNRAS.180..177B,
1980A&A....82..249B, 2006MNRAS.373...13C, 2005ApJ...626..465R} and
metallicities \citep[see, eg.,][]{2005ApJ...622.1102F,
2007MNRAS.378.1141G, 2004A&A...415.1153S}. Instead of using the
spectroscopic value for log$g$, we used the parameter $a/R_{\star}$
that is closely related to the mean stellar density and it is provided
by the transit light curve fit. The quantity $a/R_{\star}$ can be
obtained from the models using the following expression:
\begin{equation}
a/R_{\star} = \left(\frac{G}{4\pi^{2}}\right)^{1/3} \ \frac{P^{2/3}}{R_{\star}} \ \left(M_{\star} + M_{p}\right)^{1/3}.
\end{equation}
The mass of the planet is not known a priori, but $a/R_{\star}$ can be
estimated using the value of $M_{p}$ derived by
\citet{2007ApJ...671.2115B}. At the end of the procedure to be
explained next, a new value of $M_{p}$ is obtained, which is used to
repeat the process until convergence.

\begin{figure}
\plotone{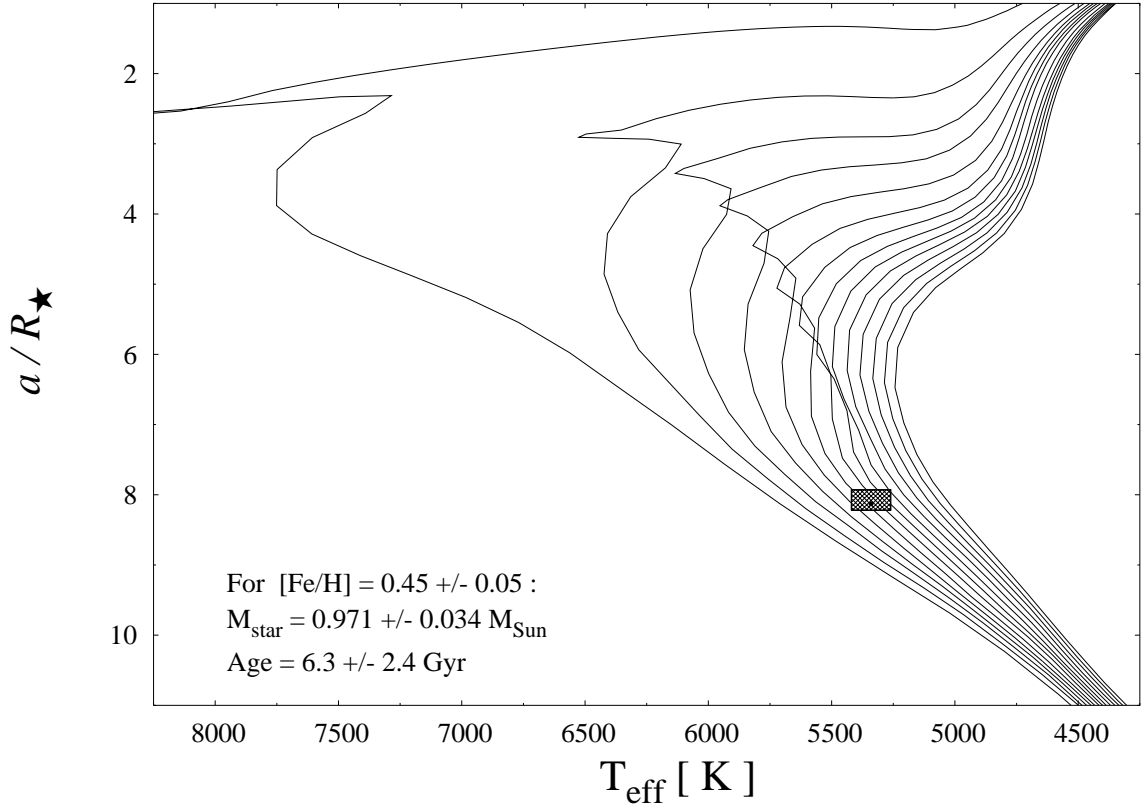}
\caption{
Yonsei-Yale stellar isochrones for the modeling of the primary star,
corresponding to ages of 0.6 Gyr, 1.6 Gyr, 2.6 Gyr, etc. (left to
right), and the chemical composition indicated.
The observed transit parameter $a/R_{\star}$, which is closely
related to the mean stellar density, was used as an indicator of
luminosity.The black dot and the filled box around it indicate the
best values and uncertainties for $a/R_{\star}$ and $T_{\rm eff}$.
\label{iso_fit}}
\end{figure}

The stellar evolution models used were those from the Yonsei-Yale
series \citep{2001ApJS..136..417Y, 2004ApJS..155..667D}. These
isochrones were interpolated to a fine grid in metallicity and age and
compared point by point with the measured values of $T_{\rm eff}$ and
$a/R_{\star}$. Each point on the isochrones that was consistent with
$\left[ {\rm Fe/H}\right]$, $T_{\rm eff}$ and $a/R_{\star}$ within
their errors, was recorded and a likelihood given by exp$\left(
  -\chi^2_{\star} / 2\right)$ was calculated, where
\begin{equation}
\chi^2_{\star} = \left(\frac{\Delta \left[{\rm Fe}/{\rm H}\right]}{\sigma_{\left[{\rm Fe}/{\rm H}\right]}}\right)^{2} \ + \ \left(\frac{\Delta T_{\rm eff}}{\sigma_{T_{\rm eff}}} \right)^{2} \ + \ \left(\frac{\Delta \left( a/R_{\star}\right)}{\sigma_{a/R_{\star}}} \right)^{2} 
\end{equation}
In this expression the $\Delta$ symbols represent the difference
between the observed and model values for each quantity. The best fit
value for each stellar parameter was obtained by adding all matches,
weighted by their corresponding normalized likelihood.  We did not
account for the varying density of stars on each isochrone, as the
effect is generally small in the case of solar-type stars. The adopted
errors for the fitted parameters (mass and age) come from their range
among the accepted points on the isochrones. Figure \ref{iso_fit}
illustrates the location of the star in a diagram of $a/R_{\star}$
vs.\ effective temperature, similar to an H-R diagram, compared with
isochrones from the Yale-Yonsei series.

With $M_{\star}$ known, we obtained $M_{p}$ and $a$ by iteration of
Newton's modified version of Kepler's third law and the mass function
of the system:
\begin{equation}
\label{kepler_3rd}
a^{3} = \frac{G}{4 \pi^{2}}\left(M_{\star} + M_{p}\right) P^2
\end{equation}
\begin{equation}
\label{mass_func}
M_{p} = \left(\frac{P}{2 \pi G}\right)^{1/3} \ \frac{K_{\star}}{\sin i}\left(M_{\star} + M_{p}\right)^{2/3}
\end{equation}
The value of $a$ in combination with $a/R_{\star}$ and
$R_{p}/R_{\star}$ allowed us to obtain consistent values for
$R_{\star}$ and $R_{p}$.

To estimate the errors for $M_{p}$, $R_{\star}$ and $R_{p}$, we used
the MCMC chains generated in the course of modeling the transiting
light curves. For each element of the chain a solution was calculated
using random Gaussian values for $P$ and $K_{\star}$ (using the
observed values and their uncertainties as the center and standard
deviation of the Gaussian distributions).  In this way we obtained a
probability distribution for $M_{p}$, $R_{\star}$, and $R_{p}$, from
which we extracted the median and 68.3\% confidence intervals and
adopted them as best values and errors, respectively.

It is important to note that our procedure places complete trust in
the Yonsei-Yale isochrones. It assumes they are exactly correct.  Any
systematic errors in the isochrones are not accounted for in our error
bars. As this star is very similar to the Sun, we expect these errors
to be small.

\section{Results}
\label{results}

Table \ref{tab_params} gives all the measured planetary and stellar
properties, together with the results for the light-curve
parameters. The labels, explained in the caption, clarify which
quantities were obtained from the literature, which were determined
independently by our analysis, which quantities are functions of those
independent parameters, and which quantities depend on our stellar
isochrone analysis.

\subsection{Planetary and stellar Properties}

The light curve best-fit parameters $a/R_{\star}$, $R_{p}/R_{\star}$
and $b$ presented in this work are consistent with those obtained by
\citet{2007ApJ...671.2115B} in the discovery paper of XO-2b, with
similar error bars for all of them. The precision of these parameters
is dominated by the first and third light curves, which have the
highest signal-to-noise ratio. For our stellar isochrone analysis, we
used the values for $T_{\rm eff}$ and [Fe/H] from the discovery paper
about XO-2b, together with our value for $a/R_{\star}$. We obtained a
stellar mass value ($M_{\star}=0.971 \pm 0.043$) almost identical to
the one found by \citet{2007ApJ...671.2115B}. In our case, the result
does not depend on the assumed distance to the star, but only on
$T_{\rm eff}$ and $\left[{\rm Fe/H}\right]$. We derived the planetary
mass and radius plus the stellar radius directly from the stellar
mass. None of these parameters show any significant discrepancy from
those listed in the discovery paper. Our error bars are similar as
well, but in this work we took into account the red noise in the light
curves and adopted larger errors for $T_{\rm eff}$ and $\left[{\rm
Fe/H}\right]$, and in this sense our error estimates are more
conservative. In the case of the stellar radius $R_{\star}$, the value
obtained directly from the isochrone modeling and the one one derived
from the stellar mass are identical, but the error bars on the latter
are 25\% smaller. Our stellar surface gravity estimate ($log \
g_{\star} = 4.45 \pm 0.01$) is independent from the one obtained by
\citet{2007ApJ...671.2115B}. Both estimates of the surface gravity are
in agreement, and our result based on $a/R_\star$ is more
precise. This illustrates the power of transit light curves to pin
down the stellar mean density, as also exhibited for the XO-3 system
\citep{2008ApJ...683.1076W}. 

As mentioned earlier, we determined the stellar properties following
the procedure and considerations of \citet{2008ApJ...677.1324T},
therefore, we expected our results for XO-2b to be in good agreement
with theirs. This is actually the case, given that our analysis of
the transit light curves didn't deliver parameters significantly
different to the ones obtained by \citet{2007ApJ...671.2115B}.

We found that the mass and radius for XO-2b are in good agreement with
those predicted by \citet{2007ApJ...659.1661F} for a planet orbiting a
solar-like star. Interpolation of the tabulated results presented on
their work for the appropriate values of the XO-2 system ($M_{p}=0.57
\mjup$, $a=0.037 \ AU$) gives a theoretical planetary radius of $1.0
\rjup$.

\subsection{Transit Timings}

Using our 6 transit times, we computed an ephemeris independent of
that of Burke et al.~(2007), by fitting a linear function of transit
epoch $E$,
\begin{equation}
\label{new_ephem}
T_{c}(E) = T_{0} + E P.
\end{equation}
The result is $T_{0}\ =\ 2,454,466.88514 \pm 0.00019$ [HJD] and $P\ =\
2.615819 \pm 0.000014$ days. The fit has $\chi^2=5.76$ with 4 degrees
of freedom ($N_{\rm dof}=4$), indicating an acceptable fit (i.e.\
$\chi^2$ is within $\sqrt{2N_{\rm dof}}$ of $N_{\rm dof}$). However,
the fitted period differs by 2.5$\sigma$ (3.3~s) from the value
$2.615857\pm 0.000005$~d that was determined by Burke et al.~(2007).

It is not clear how to interpret this discrepancy. It may be a genuine
period variation produced by an additional body in the XO-2
system. However it seems at least as likely that the uncertainties in
one or a few of the transit times have been underestimated. Our
methods and cross-checks on the errors have already been described. As
for Burke et al.~(2007), one of their mid-transit times was based on
observations with a 1.8m telescope, and has a quoted precision that is
better than any of our 6 measurements ($2,454,147.74902 \pm
0.00020$). The other 11 light curves came from observations with
smaller telescopes. For those 11, Burke et al.~(2007) estimated an
error of 3~min, based on the scatter among 3 independent measurements
of one particular transit in March~2007. The relative quality of the
individual light curves, which were gathered by different observers on
different nights, was not taken into account.

One might wonder if the discrepancy can be attributed to any
particular data point. If we use only the most precise time of Burke
et al.~(2007) in combination with our 6 times, fitting a linear
ephemeris gives $\chi^2=15.9$ and $N_{\rm dof}=5$, which is still
unacceptable. If we omit the most precise time and instead use the
other 11 times from Burke et al.~(2007), adopting 3~min errors as they
did, we find $\chi^2=20.9$ and $N_{\rm dof}=15$, right on the margin
of acceptability. If we use all of the times from Burke et al.~(2007)
and omit our last 2 times, which were derived from the noisiest light
curves, we find $\chi^2=15.6$ and $N_{\rm dof}=14$, a satisfactory
fit. (Omitting either of our last two times individually does not
result in a satisfactory fit.) No further conclusions can be drawn
until more times have been measured, over a longer time baseline.

For planning purposes, we recomputed the ephemeris based on all the
available transit times (6 from this work, and 12 from Burke et
al.~2007). The result is $T_{0}=2,454,466.88467\pm 0.00013$~[HJD] and
$P=2.6158640\pm 0.0000016$~d, with $\chi^2=27.6$ and $N_{\rm dof}=16$.
Observers who are using these values to plan future observations may
wish to inflate the error bars by $\sqrt{\chi^2/N_{\rm dof}} = 1.31$
to account for possible systematic effects due to either genuine
period variations or underestimated measurement errors.

A more detailed study of possible short-term transiting timing
variations caused by additional orbiting bodies would require the
observation of a larger number of consecutive transits, as noted by
\citet{2005Sci...307.1288H} and tested by
\citet{2008ApJ...682..593M}. Any constraint derived from the present
data would be of limited interest and beyond of the scope of this
paper.

\section{Summary}
\label{summ}

We have presented new photometric observations of six transits of the
exoplanet XO-2b. The analysis of the data independently confirms the
previously calculated properties of this planet, leaving little doubt
about the orbit and geometric properties of the XO-2 system. We found
that the mass and radius for XO-2b are in good agreement with standard
models of gas giant planets. The six new transit timings yield a
refined transit ephemeris and set an updated and precise reference for
future searches of secondary bodies orbiting this star.

\acknowledgments{This work was partly supported by NASA Origins Grant
  No.~NNX09AB33G, and by Grant No.\ 2006234 from the United
  States-Israel Binational Science Foundation (BSF).  KeplerCam was
  developed with partial support from the Kepler Mission under NASA
  Cooperative Agreement NCC2-1390 and the Keplercam observations
  described in this paper were partly supported by grants from the
  Kepler Mission to SAO and PSI.}

\bibliography{adssample}

\begin{thebibliography}{32}
\expandafter\ifx\csname natexlab\endcsname\relax\def\natexlab#1{#1}\fi

\bibitem[{{Agol} {et~al.}(2005){Agol}, {Steffen}, {Sari}, \&
  {Clarkson}}]{2005MNRAS.359..567A}
{Agol}, E., {Steffen}, J., {Sari}, R., \& {Clarkson}, W. 2005, \mnras, 359, 567

\bibitem[{{Beatty} {et~al.}(2007){Beatty}, {Fern{\'a}ndez}, {Latham}, {Bakos},
  {Kov{\'a}cs}, {Noyes}, {Stefanik}, {Torres}, {Everett}, \&
  {Hergenrother}}]{2007ApJ...663..573B}
{Beatty}, T.~G., {Fern{\'a}ndez}, J.~M., {Latham}, D.~W., {Bakos}, G.~{\'A}.,
  {Kov{\'a}cs}, G., {Noyes}, R.~W., {Stefanik}, R.~P., {Torres}, G., {Everett},
  M.~E., \& {Hergenrother}, C.~W. 2007, \apj, 663, 573

\bibitem[{{Blackwell} {et~al.}(1980){Blackwell}, {Petford}, \&
  {Shallis}}]{1980A&A....82..249B}
{Blackwell}, D.~E., {Petford}, A.~D., \& {Shallis}, M.~J. 1980, \aap, 82, 249

\bibitem[{{Blackwell} \& {Shallis}(1977)}]{1977MNRAS.180..177B}
{Blackwell}, D.~E., \& {Shallis}, M.~J. 1977, \mnras, 180, 177

\bibitem[{{Bouchy} {et~al.}(2005){Bouchy}, {Pont}, {Melo}, {Santos}, {Mayor},
  {Queloz}, \& {Udry}}]{2005A&A...431.1105B}
{Bouchy}, F., {Pont}, F., {Melo}, C., {Santos}, N.~C., {Mayor}, M., {Queloz},
  D., \& {Udry}, S. 2005, \aap, 431, 1105

\bibitem[{{Brosch} {et~al.}(2008){Brosch}, {Polishook}, {Shporer}, {Kaspi},
  {Berwald}, \& {Manulis}}]{2008Ap&SS.314..163B}
{Brosch}, N., {Polishook}, D., {Shporer}, A., {Kaspi}, S., {Berwald}, A., \&
  {Manulis}, I. 2008, \apss, 314, 163

\bibitem[{{Burke} {et~al.}(2007){Burke}, {McCullough}, {Valenti},
  {Johns-Krull}, {Janes}, {Heasley}, {Summers}, {Stys}, {Bissinger}, {Fleenor},
  {Foote}, {Garc{\'{\i}}a-Melendo}, {Gary}, {Howell}, {Mallia}, {Masi},
  {Taylor}, \& {Vanmunster}}]{2007ApJ...671.2115B}
{Burke}, C.~J., {McCullough}, P.~R., {Valenti}, J.~A., {Johns-Krull}, C.~M.,
  {Janes}, K.~A., {Heasley}, J.~N., {Summers}, F.~J., {Stys}, J.~E.,
  {Bissinger}, R., {Fleenor}, M.~L., {Foote}, C.~N., {Garc{\'{\i}}a-Melendo},
  E., {Gary}, B.~L., {Howell}, P.~J., {Mallia}, F., {Masi}, G., {Taylor}, B.,
  \& {Vanmunster}, T. 2007, \apj, 671, 2115

\bibitem[{{Casagrande} {et~al.}(2006){Casagrande}, {Portinari}, \&
  {Flynn}}]{2006MNRAS.373...13C}
{Casagrande}, L., {Portinari}, L., \& {Flynn}, C. 2006, \mnras, 373, 13

\bibitem[{{Charbonneau} {et~al.}(2007){Charbonneau}, {Brown}, {Burrows}, \&
  {Laughlin}}]{2007prpl.conf..701C}
{Charbonneau}, D., {Brown}, T.~M., {Burrows}, A., \& {Laughlin}, G. 2007, in
  Protostars and Planets V, ed. B.~{Reipurth}, D.~{Jewitt}, \& K.~{Keil},
  701--716

\bibitem[{{Claret}(2004)}]{2004A&A...428.1001C}
{Claret}, A. 2004, \aap, 428, 1001

\bibitem[{{Demarque} {et~al.}(2004){Demarque}, {Woo}, {Kim}, \&
  {Yi}}]{2004ApJS..155..667D}
{Demarque}, P., {Woo}, J.-H., {Kim}, Y.-C., \& {Yi}, S.~K. 2004, \apjs, 155,
  667

\bibitem[{{Fischer} \& {Valenti}(2005)}]{2005ApJ...622.1102F}
{Fischer}, D.~A., \& {Valenti}, J. 2005, \apj, 622, 1102

\bibitem[{{Ford}(2005)}]{2005AJ....129.1706F}
{Ford}, E.~B. 2005, \aj, 129, 1706

\bibitem[{{Fortney} {et~al.}(2007){Fortney}, {Marley}, \&
  {Barnes}}]{2007ApJ...659.1661F}
{Fortney}, J.~J., {Marley}, M.~S., \& {Barnes}, J.~W. 2007, \apj, 659, 1661

\bibitem[{{Gonzalez} \& {Laws}(2007)}]{2007MNRAS.378.1141G}
{Gonzalez}, G., \& {Laws}, C. 2007, \mnras, 378, 1141

\bibitem[{{Holman} \& {Murray}(2005)}]{2005Sci...307.1288H}
{Holman}, M.~J., \& {Murray}, N.~W. 2005, Science, 307, 1288

\bibitem[{{Holman} {et~al.}(2006){Holman}, {Winn}, {Latham}, {O'Donovan},
  {Charbonneau}, {Bakos}, {Esquerdo}, {Hergenrother}, {Everett}, \&
  {P{\'a}l}}]{2006ApJ...652.1715H}
{Holman}, M.~J., {Winn}, J.~N., {Latham}, D.~W., {O'Donovan}, F.~T.,
  {Charbonneau}, D., {Bakos}, G.~A., {Esquerdo}, G.~A., {Hergenrother}, C.,
  {Everett}, M.~E., \& {P{\'a}l}, A. 2006, \apj, 652, 1715

\bibitem[{{Holman} {et~al.}(2007){Holman}, {Winn}, {Latham}, {O'Donovan},
  {Charbonneau}, {Torres}, {Sozzetti}, {Fernandez}, \&
  {Everett}}]{2007ApJ...664.1185H}
{Holman}, M.~J., {Winn}, J.~N., {Latham}, D.~W., {O'Donovan}, F.~T.,
  {Charbonneau}, D., {Torres}, G., {Sozzetti}, A., {Fernandez}, J., \&
  {Everett}, M.~E. 2007, \apj, 664, 1185

\bibitem[{{Johnson} {et~al.}(2008){Johnson}, {Winn}, {Cabrera}, \&
  {Carter}}]{2008arXiv0812.0029J}
{Johnson}, J.~A., {Winn}, J.~N., {Cabrera}, N.~E., \& {Carter}, J.~A. 2008,
  ArXiv e-prints

\bibitem[{{Mandel} \& {Agol}(2002)}]{2002ApJ...580L.171M}
{Mandel}, K., \& {Agol}, E. 2002, \apjl, 580, L171

\bibitem[{{McCullough} {et~al.}(2006){McCullough}, {Stys}, {Valenti},
  {Johns-Krull}, {Janes}, {Heasley}, {Bye}, {Dodd}, {Fleming}, {Pinnick},
  {Bissinger}, {Gary}, {Howell}, \& {Vanmunster}}]{2006ApJ...648.1228M}
{McCullough}, P.~R., {Stys}, J.~E., {Valenti}, J.~A., {Johns-Krull}, C.~M.,
  {Janes}, K.~A., {Heasley}, J.~N., {Bye}, B.~A., {Dodd}, C., {Fleming}, S.~W.,
  {Pinnick}, A., {Bissinger}, R., {Gary}, B.~L., {Howell}, P.~J., \&
  {Vanmunster}, T. 2006, \apj, 648, 1228

\bibitem[{{Miller-Ricci} {et~al.}(2008){Miller-Ricci}, {Rowe}, {Sasselov},
  {Matthews}, {Kuschnig}, {Croll}, {Guenther}, {Moffat}, {Rucinski}, {Walker},
  \& {Weiss}}]{2008ApJ...682..593M}
{Miller-Ricci}, E., {Rowe}, J.~F., {Sasselov}, D., {Matthews}, J.~M.,
  {Kuschnig}, R., {Croll}, B., {Guenther}, D.~B., {Moffat}, A.~F.~J.,
  {Rucinski}, S.~M., {Walker}, G.~A.~H., \& {Weiss}, W.~W. 2008, \apj, 682, 593

\bibitem[{{Press} {et~al.}(1992){Press}, {Teukolsky}, {Vetterling}, \&
  {Flannery}}]{1992nrfa.book.....P}
{Press}, W.~H., {Teukolsky}, S.~A., {Vetterling}, W.~T., \& {Flannery}, B.~P.
  1992, {Numerical recipes in FORTRAN. The art of scientific computing}
  (Cambridge: University Press, |c1992, 2nd ed.)

\bibitem[{{Ram{\'{\i}}rez} \& {Mel{\'e}ndez}(2005)}]{2005ApJ...626..465R}
{Ram{\'{\i}}rez}, I., \& {Mel{\'e}ndez}, J. 2005, \apj, 626, 465

\bibitem[{{Santos} {et~al.}(2004){Santos}, {Israelian}, \&
  {Mayor}}]{2004A&A...415.1153S}
{Santos}, N.~C., {Israelian}, G., \& {Mayor}, M. 2004, \aap, 415, 1153

\bibitem[{{Seager} \& {Mall{\'e}n-Ornelas}(2003)}]{2003ApJ...585.1038S}
{Seager}, S., \& {Mall{\'e}n-Ornelas}, G. 2003, \apj, 585, 1038

\bibitem[{{Southworth}(2008)}]{2008MNRAS.386.1644S}
{Southworth}, J. 2008, \mnras, 386, 1644

\bibitem[{{Southworth} {et~al.}(2004){Southworth}, {Zucker}, {Maxted}, \&
  {Smalley}}]{2004MNRAS.355..986S}
{Southworth}, J., {Zucker}, S., {Maxted}, P.~F.~L., \& {Smalley}, B. 2004,
  \mnras, 355, 986

\bibitem[{{Sozzetti} {et~al.}(2007){Sozzetti}, {Torres}, {Charbonneau},
  {Latham}, {Holman}, {Winn}, {Laird}, \& {O'Donovan}}]{2007ApJ...664.1190S}
{Sozzetti}, A., {Torres}, G., {Charbonneau}, D., {Latham}, D.~W., {Holman},
  M.~J., {Winn}, J.~N., {Laird}, J.~B., \& {O'Donovan}, F.~T. 2007, \apj, 664,
  1190

\bibitem[{{Torres} {et~al.}(2008){Torres}, {Winn}, \&
  {Holman}}]{2008ApJ...677.1324T}
{Torres}, G., {Winn}, J.~N., \& {Holman}, M.~J. 2008, \apj, 677, 1324

\bibitem[{{Winn} {et~al.}(2008){Winn}, {Holman}, {Torres}, {McCullough},
  {Johns-Krull}, {Latham}, {Shporer}, {Mazeh}, {Garcia-Melendo}, {Foote},
  {Esquerdo}, \& {Everett}}]{2008ApJ...683.1076W}
{Winn}, J.~N., {Holman}, M.~J., {Torres}, G., {McCullough}, P., {Johns-Krull},
  C., {Latham}, D.~W., {Shporer}, A., {Mazeh}, T., {Garcia-Melendo}, E.,
  {Foote}, C., {Esquerdo}, G., \& {Everett}, M. 2008, \apj, 683, 1076

\bibitem[{{Yi} {et~al.}(2001){Yi}, {Demarque}, {Kim}, {Lee}, {Ree}, {Lejeune},
  \& {Barnes}}]{2001ApJS..136..417Y}
{Yi}, S., {Demarque}, P., {Kim}, Y.-C., {Lee}, Y.-W., {Ree}, C.~H., {Lejeune},
  T., \& {Barnes}, S. 2001, \apjs, 136, 417

\end{thebibliography}

\clearpage
\begin{deluxetable}{lcccccc}
\tabletypesize{\tiny}
\tablewidth{0pt}
\tablecaption{\sc Photometric Observations Summary \label{tab_phot}}
\tablehead{}
\startdata
UT Date               & Jan 01 2008   & Jan 14 2008   &  Feb 12 2008   &  Feb 25 2008  &  Mar 04 2008   & Mar 06 2008      \\
Epoch$^1$             & 0             & 5             &  16            &  21           &  24            & 25              \\
Observatory           & FLWO          & FLWO          &  FLWO          &  FLWO         &  FLWO          & Wise             \\
Band                  & Sloan $z$     & Sloan $z$     &  Sloan $z$     &  Sloan $z$    &  Sloan $z$     & clear            \\
Exposure [sec]         & $ 30        $ & $ 25        $ & $ 25         $ & $ 30         $& $ 25         $ & $ 9           $  \\
FWHM$_{\rm median}$ [arcsec] & $ 1.6    $ & $ 1.5       $ & $ 1.2        $ & $ 1.4        $& $ 2.9        $ & $ 1.3         $  \\
RMS     [rel.\ flux]   & $ 0.0013    $ & $ 0.0013    $ & $ 0.0017     $ & $ 0.0027     $& $ 0.0026     $ & $ 0.0036      $  \\
\enddata
\tablecomments{(1)Number of cycles elapsed since the initial transit studied in this work }
\end{deluxetable}

\clearpage
\begin{deluxetable}{ccc}
\tabletypesize{\footnotesize}
\tablewidth{0pc}
\tablecaption{\sc Transit Light Curve \label{sample_lc}}
\tablehead{\multicolumn{2}{c}{Jan 01 2008 UT} \\
\hline\\
\colhead{HJD}             &
\colhead{Relative Flux}
}
\startdata
 2454466.78551   &  0.99807 \\
 2454466.78602   &  1.00079 \\
 2454466.78654   &  1.00180 \\
 2454466.78705   &  0.99955 \\
\enddata
\tablecomments{
This is a sample entry of a full light curve. The complete versions are given on-line.}
\end{deluxetable}

\clearpage
\begin{deluxetable}{lcccc}
\tablewidth{0pt}
\tablecaption{\sc Transit Timings and  System Parameters for XO-2 \label{tab_params}}
\tablehead{
\colhead{Parameter}&
\colhead{}&
\colhead{Value}&
\colhead{68.3\% Conf. Limits}&
\colhead{Comment}}
\startdata
$T_{c}$ (0)     & [HJD]              & $ 2,454,466.88512 $ & $ 0.00021                $ & A \\
$T_{c}$ (5)     & [HJD]              & $ 2,454,479.96393 $ & $ 0.00039                $ & A \\
$T_{c}$ (16)     & [HJD]              & $ 2,454,508.73864 $ & $ 0.00026                $ & A \\
$T_{c}$ (21)     & [HJD]              & $ 2,454,521.81778 $ & $ 0.00072                $ & A \\
$T_{c}$ (24)     & [HJD]              & $ 2,454,529.66433 $ & $ 0.00043                $ & A \\
$T_{c}$ (25)     & [HJD]              & $ 2,454,532.27978 $ & $ 0.00074                $ & A \\
\\
$P$              & [days]             & $ 2.6158640     $ & $ 0.0000016              $ & B \\
$T_{0}$          & [HJD]              & $ 2,454,466.88467 $ & $ 0.00012                $ & B \\
\\
$R_{p}/R_{\star}$ &                    & $ 0.10485       $ & $ +0.00070, -0.00062     $ & A \\
$a/R_{\star}$     &                    & $ 8.13          $ & $ +0.09, -0.20           $ & A \\
$b$               &                    & $ 0.16          $ & $ 0.11                   $ & A \\
$u_{1}$           &                    & $ 0.250         $ & $ 0.026                  $ & A \\
$u_{2}$           &                    & $ 0.296         $ & $ 0.013                  $ & A \\
\\
$\rho_{star}$     & [$g \ cm^{-3}$]    & $ 1.484         $ & $ +0.051, -0.104         $ & C \\
$log \ g_{p}$     & [$cm \ s^{-2}$]  & $ 3.147         $ & $ +0.043, -0.048         $ & C \\
\\
$M_{\star}$       & [$\msun$]          & $ 0.971         $ & $ 0.034                  $ & D \\
$age_{\star}$     & [Gyr]              & $ 6.3           $ & $ 2.4                    $ & D \\
\\
$R_{\star}$       & [$\rsun$]          & $ 0.976         $ & $ +0.024, -0.016         $ & E \\
$log \ g_{\star}$ & [$cm \ seg^{-2}$]  & $ 4.448         $ & $ +0.011, -0.021         $ & E \\
$M_{p}$           & [$\mjup$]          & $ 0.565         $ & $ 0.054                  $ & E \\
$R_{p}$           & [$\rjup$]          & $ 0.996         $ & $ +0.031, -0.018         $ & E \\
$a$               & [AU]               & $ 0.0368        $ & $ 0.0004                 $ & E \\
\enddata
\tablecomments{
(A) obtained independently. (B) derived using transit timings from
this work and from \citet{2007ApJ...671.2115B}. (C) calculated using
$K_{\star}$ from \citet{2007ApJ...671.2115B}.(D) derived from
isochrone modeling using values for $T_{\rm eff}$ and
$\left[{\rm Fe/H}\right]$ from \citet{2007ApJ...671.2115B}.  (E) function of
A, C and D }
\end{deluxetable}

\clearpage

\begin{figure}
\plotone{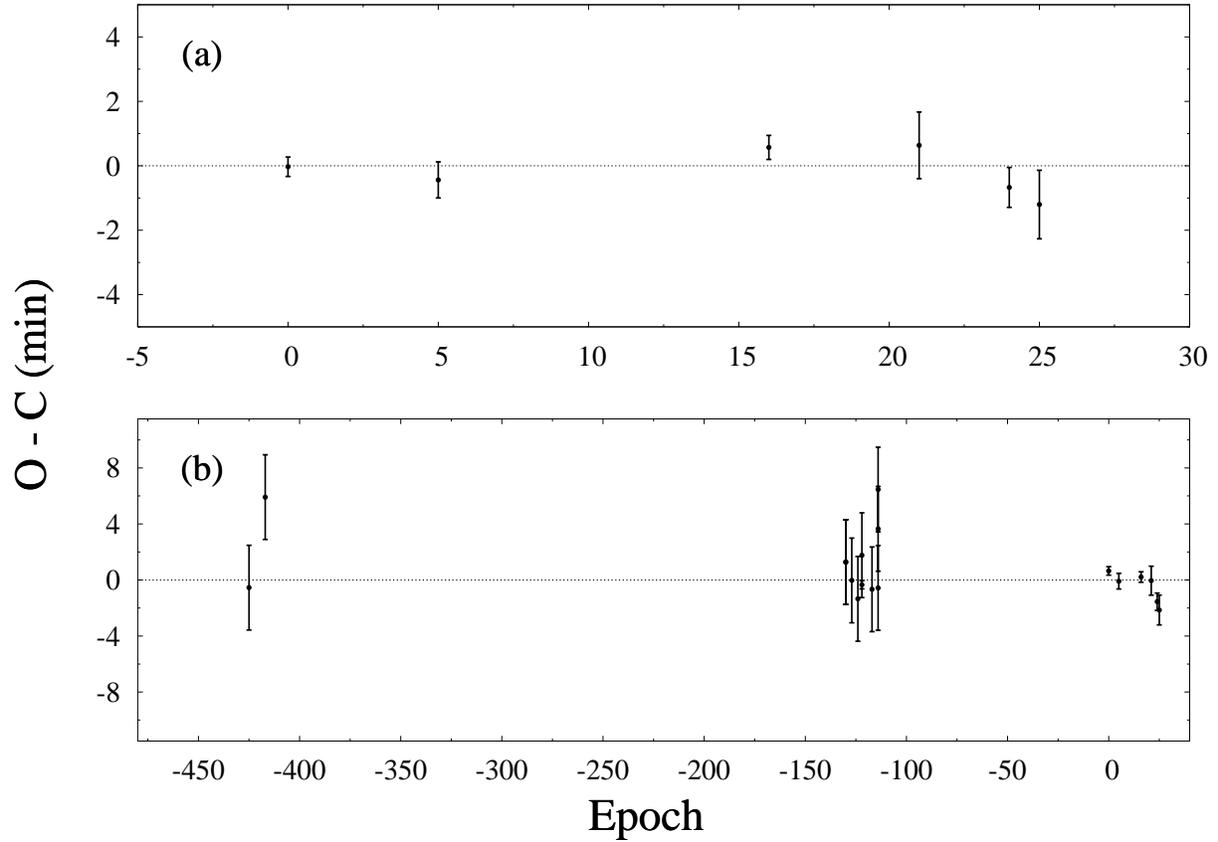}
\caption{Differences between observed and calculated transit times. On
  panel (a), the 6 TLC transit timings are shown. The time residuals
  are calculated from the ephemeris obtained using TLC transits
  only. Panel (b) shows the time residuals of all the transits
  reported on the discovery paper plus those presented in this
  work. The time residuals are calculated from the ephemeris obtained
  using all the available timings.\label{ttv}}
\end{figure}

\end{document}